\documentclass[useAMS,usenatbib]{mn2e}
\usepackage{graphicx}
\usepackage{array,booktabs,tabularx}

\title[Wind-Driven X-Ray Spectral Variability of NGC 1365]{Long Term Wind-Driven X-Ray Spectral Variability of NGC 1365 with {\it
Swift}}

\author[S.D. Connolly, I.M.  McHardy and T. Dwelly]{S. D. Connolly$^{1}$\thanks{E-mail: sdc1g08@soton.ac.uk}, I.M.
M\parbox[b][3.4mm][t]{2.0mm}{c}Hardy$^{1}$, T.
Dwelly$^{2,1}$ \\
$^{1}$University of Southampton, Highfield, Southampton, SO17 1BJ, UK\\
$^{2}$Max-Planck-Institut f\"{u}r Extraterrestrische Physik, Giessenbachstrasse 1, 85748, Garching, DE}

\begin{document}
\date{Accepted 2014 March 17. Received 2014 March 17; in original form 2014 January 06}
\pagerange{\pageref{firstpage}--\pageref{lastpage}} \pubyear{2014}

\maketitle

\label{firstpage}


\begin{abstract}

We present long-term (months-years) X-ray spectral variability of the
Seyfert 1.8 galaxy NGC 1365 as observed by {\it Swift}, which provides
well sampled observations over a much longer timescale (6 years) and a
much larger flux range than is afforded by other observatories. At
very low luminosities the spectrum is very soft, becoming rapidly
harder as the luminosity increases and then, above a particular
luminosity, softening again.  At a given flux level, the scatter
in hardness ratio is not very large, meaning that the spectral shape is
largely determined by the luminosity.  The spectra were therefore summed in
luminosity bins and fitted with a variety of models. The best fitting model
consists of two power laws, one unabsorbed and another, more luminous,
which is absorbed. In this model, we find a range of intrinsic 0.5-10.0 keV luminosities of approximately $1.1 - 3.5$ ergs s$^{-1}$,
and a very large range of absorbing columns, of approximately $10^{22} - 10^{24}$ cm$^{-2}$. 
Interestingly, we find that the absorbing column
decreases with increasing luminosity, but that this result is not due
to changes in ionisation. We suggest that these observations might be
interpreted in terms of a wind model in which the launch radius varies as a 
function of ionising flux and disc temperature and therefore moves out with increasing
accretion rate, i.e. increasing X-ray luminosity. Thus, depending
on the inclination angle of the disc relative to the observer, the absorbing column may decrease
as the accretion rate goes up. The weaker, unabsorbed, component may
be a scattered component from the wind. 

\end{abstract}


\section{Introduction}
\label{intro}

X-ray spectral observations have shown that variability in the column of absorbing material between the X-ray source and the observer is present
in a number of Seyfert galaxies \citep{risaliti02}. The detection of variable absorption on a timescales of hours has indicated that the absorbing material must be close
to the nucleus, at a distance similar to that of the Broad Emission Line Region   \citep[e.g.][]{lamer,elvis04,puccetti}, with claims that complete occultations by Broad
Line Region clouds have been observed on timescales of days \citep{risaliti07a}.

NGC 1365 is a nearby Seyfert 1.8 galaxy \citep{maiolino95} which displays a large amount of X-ray spectral variability \citep{risaliti09} on timescales of hours
to years \citep{brenneman}. These variations have been interpreted as the spectrum changing from being `transmission dominated' to `reflection
dominated'. When the spectrum is `transmission dominated' \citep[e.g.][]{risaliti00} the absorbing material is Compton thin and the transmitted component dominates
the spectrum; when the spectrum is `reflection dominated'  \citep[e.g.][]{iyomoyo} the absorbing material is Compton thick, meaning the majority of direct emission is
absorbed and reflected emission dominates the spectrum \citep{risaliti07b,matt}. 

A number of absorption and emission lines have been seen in the spectrum. A strong Fe fluorescence emission line is present at 6.4 keV, together with a group of Fe
absorption lines between 6.7 and 8.3 keV, attributed to FeXXV and FeXXVI K$\alpha$ and K$\beta$ transitions. The measured velocities of these lines has lead to
speculation that they originate from a highly-ionised, high-velocity outflow from NGC 1365 \citep{risaliti05a}.

Although there have been many previous X-ray spectral studies of NGC 1365, these studies have all concentrated either on detailed analysis of a single epoch spectrum or
on analysis of a small number of spectra taken over a relatively short timescale (hours or days). By contrast, here we study 190 {\it Swift} spectra taken over a period
of six years. Whilst {\it Swift} does not provide spectral resolution as high as that used in most previous short-time X-ray spectral studies, e.g. with {\it XMM-Newton}
or {\it Suzaku}, the {\it Swift} data cover a much longer time period and a far greater flux range. The {\it Swift} data therefore allow a proper investigation of
flux-related spectral variability and of long term spectral variations, over a much larger dynamic range than in previous studies.


The spectrum of NGC 1365, as with most AGN, has previously been modelled using a power law component, with an intrinsic spectral index, $\Gamma$. It is not known whether
$\Gamma$ varies or not during changes in X-ray luminosity. Many groups \citep[e.g.][]{miller08,turner07,fabian05,pounds04} assume that there is no change. 
Observations in the $2-10$ keV band generally do show some
variation, although the changes with luminosity are not large \citep[e.g.][]{sobolewska,zdziarski99}. \citet{sobolewska}, for example, who simply fit a power
law to the 2-10 keV spectra, find that the observed $\Gamma$ varies as 2.7$\dot \mathrm{m} ^{0.08}$ over similar time scales to that of our data. In reality, however,
these
measurements of $\Gamma$ are, of course, depend on other parameters which were not included in the fits, such as the reflection component and any absorption. If the
variation in the observed spectral index is interpreted in terms of the sum of a variable, steep spectrum component and a relatively constant reflection component with a
hard spectrum, the intrinsic spectral index can remain constant; in this case, when the flux of the variable steep spectrum component is low, the hard spectrum component
dominates, causing the observed spectral index to change  \citep[e.g.][]{guainazzi99,uttley99,ponti,fabian03}. Furthermore, where observations with a large spectral range
have been
made, allowing good definition of the primary continuum slope, the observed variation of $\Gamma$ with luminosity has not been large (e.g. 0.1 in NGC 4151,
\citealt{lubinski10}), 0.2 in NGC 4507, \citealt{braito}).

Theoretical Comptonisation modelling  \citep[e.g.][]{beloborodov99,coppi92} shows that the photon index can depend on the ratio of L$_{diss}$ to L$_s$ (where
L$_{diss}$ is the power dissipated in the corona during variations and L$_s$ is the input soft photon luminosity) to a low power (-0.1 for AGN). Unless there are very
large variations in these parameters, the intrinsic spectral index should therefore not change by more than a few tenths. Thus, although it is possible that a small
change in spectral index may occur over the flux range sampled by our observations, the large changes in $\Gamma$ required by the pivoting power law models are assumed
to be unlikely. In many of our models, including the model we deem most accurate, we therefore assume $\Gamma$ to be constant, as this
is likely to be a reasonable approximation.

The variability of the spectrum of NGC 1365 has previously been modelled using a partial covering model, in which a varying fraction of the X-ray source is
  obscured by absorbing material \citep[e.g.][]{risaliti09}. This model has been found to fit the data for individual, short-timescale events and is therefore also
tested here.

\section{Observations \& Data Reduction}
\label{obs}

\begin{figure*}
	\includegraphics[width = 17.5cm,clip = true, trim = 0 0 0 0]{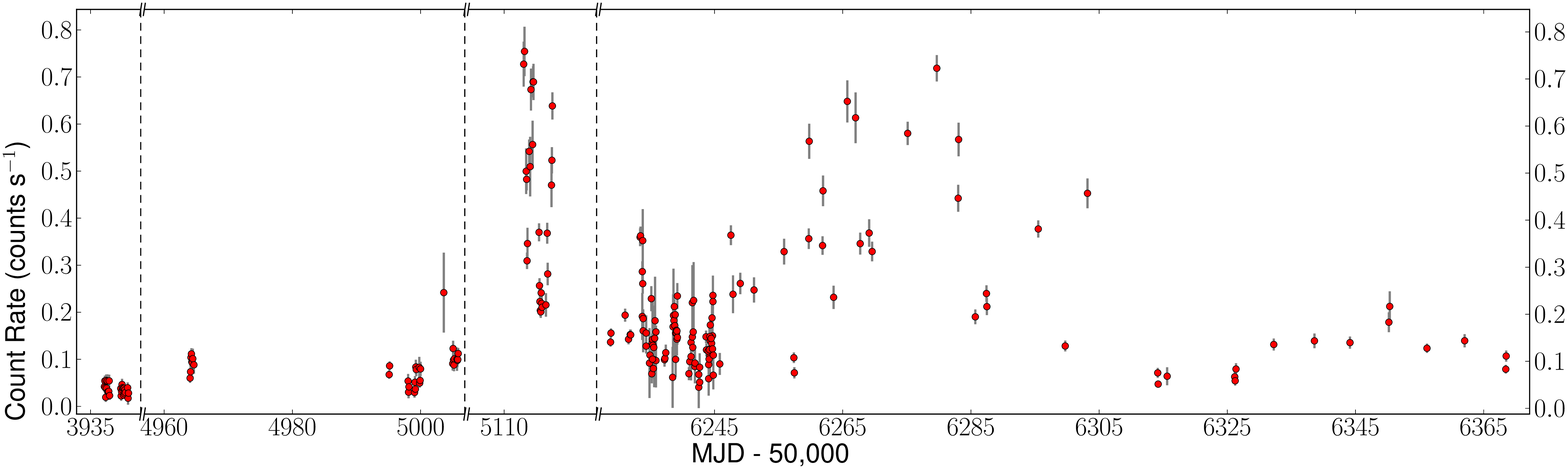}
	
	\caption{The {\it Swift} X-ray lightcurve of NGC 1365, with a broken axis where data were not taken. The section on the right shows the period during which more
intensive {\it Swift} monitoring was taking place as a result of SN2012fr.}
	\label{lc}
\end{figure*}

Observations of NGC 1365 using the {\it Swift} satellite have been carried out as part of a number of different programmes, the combined data from which has been used in
this study. The observations were performed using the {\it Swift} XRT in `photon counting mode', between 21 July 2006 and 17 March 2013. A total of 293 spectra from
individual
{\it Swift} `visits', or exposures, were used, with a total of more than $220$ kiloseconds of exposure time. Individual exposure times ranged from $<10$ seconds to
$>1900$ seconds. Data were mostly confined to three main time periods, shown in Table \ref{obstable}. More intensive monitoring took place between MJD 56220 - 56330, due
to Supernova 2012fr, which went off in NGC 1365 during this period (but which was neither bright enough nor close enough to the nucleus to affect the data used in this
study  \citep[e.g.][]{klotz,childress}). Of the 293 spectra, 103 were rejected due to either very low signal to noise or artifacts near the source,
leaving 190 usable spectra, consisting of $\sim165$ kiloseconds of exposure time and over 25000 photon counts. The raw data for all {\it Swift XRT} observations of NGC
1365 were downloaded from the HEASARC archive$^1$\footnotetext[1]{http://heasarc.gsfc.nasa.gov/cgi-bin/W3Browse/swift.pl}.

The XRT data were reduced using an automatic pipeline, fully described in \citet{fabian12} and previously used in e.g. \citet{cameron}. In each case, the reduction used
the most recent version of the standard {\it Swift} XRTPIPELINE software (versions 0.12.4 - 0.12.6). The XSELECT tool was used to extract spectra and lightcurves, using
flux-dependent source and background extraction regions which were chosen such as to reduce the background contamination at faint fluxes, and to mitigate the effects
of pile-up at high fluxes. The sensitivity of the XRT is not uniform over the field of view, due to vignetting and the presence of bad pixels and columns on the CCD; the
{\it Swift} XRTEXPOMAP and XRTMKARF tools were therefore used to generate an exposure map (including vignetting and bad pixels) and an ancillary response file (ARF) for
each visit, in order to correct for these effects. The relevant redistribution matrix file (RMF) from the {\it Swift} calibration database was also supplied in each case.
The local X-ray background was estimated and subtracted from the instrumental count rates, using the area-scaled count rate measured in a background annulus region. The
observed XRT count rates were carefully corrected for the fraction of counts lost due to bad pixels and columns, vignetting effects, and
the finite extraction aperture (including regions excised in order to mitigate pileup effects).

Fig. \ref{lc} shows a $0.5 - 10.0$ keV lightcurve of all of the {\it Swift} data over the six-year period of observation. The large flux range is readily apparent on a
range of timescales.

{\it Chandra} images show that the nucleus of NGC 1365 is embedded within a region of extended low surface brightness emission, of radius approximately 15 arcsec
(see \citealt{wang09}). The region used to determine the {\it Swift} background contribution lies outside this small region of extended emission. Thus, a contribution
from the constant extended emission within the {\it Swift} PSF will remain in our {\it Swift} spectra. To determine what that contribution is, we examined {\it Chandra}
spectra of the extended emission region. The {\it Chandra} data used were taken in December 2002, using the ACIS-S instrument (OBSID 3554). Four spectra were extracted
from circular regions, of size similar to the {\it Swift} PSF, lying close to the nucleus. The spectra were extracted from the the event files in the primary (reduced)
data set,  using the CIAO 4.6 tool `specextract'. Extraction regions of radius 2 arcsecs were used in each case.

The background spectra were taken from within the same region from which the {\it Swift} background spectra were taken, such that the resultant spectra did
not contain the background which had been subtracted from the {\it Swift} spectra. Circular extraction regions of radius 2 arcsecs were also used.


\begin{table}
 	\centering
	\begin{tabular}{ p{2.05cm} p{2.1cm} p{0.3cm}  p{0.6cm}  p{.95cm} p{0.6cm} }
	\hline
		
	OBSIDs 							& MJD range 		& N$_{obs}$ 	& N$_{visits}$ 	& T$_{tot}$ (s) & Cnts$_{tot}$	\\ \hline
	00035458001-02						& 53937.0-53940.8	& 2		& 20		& 14714		& 433		\\
	00090101001-09						& 54964.0-55117.5	& 8		& 56 		& 44576		& 8795		\\ 
	00035458003,						&			&		&		&		&		\\
	00080317001, 						& 56134.1-56368.5	& 41		& 114		& 105515	& 16227		\\ 
	00032614001-71 						&			&		&		&		&		\\ \hline
	Total							& 53937.0-56368.5	& 51		& 190		& 164805	& 25455		\\ 
	\hline
	\end{tabular}
		
	\caption{Summary of {\it Swift} observations used in this work. N$_{obs}$, N$_{visits}$ and T$_{tot}$  are the values remaining after unusable data has been
excluded.}
	\label{obstable}

\end{table}

\section{Data Analysis}

\subsection{Spectral Hardness}

\begin{figure}
	\includegraphics[width = \columnwidth,clip = true, trim = 0 0 0 0]{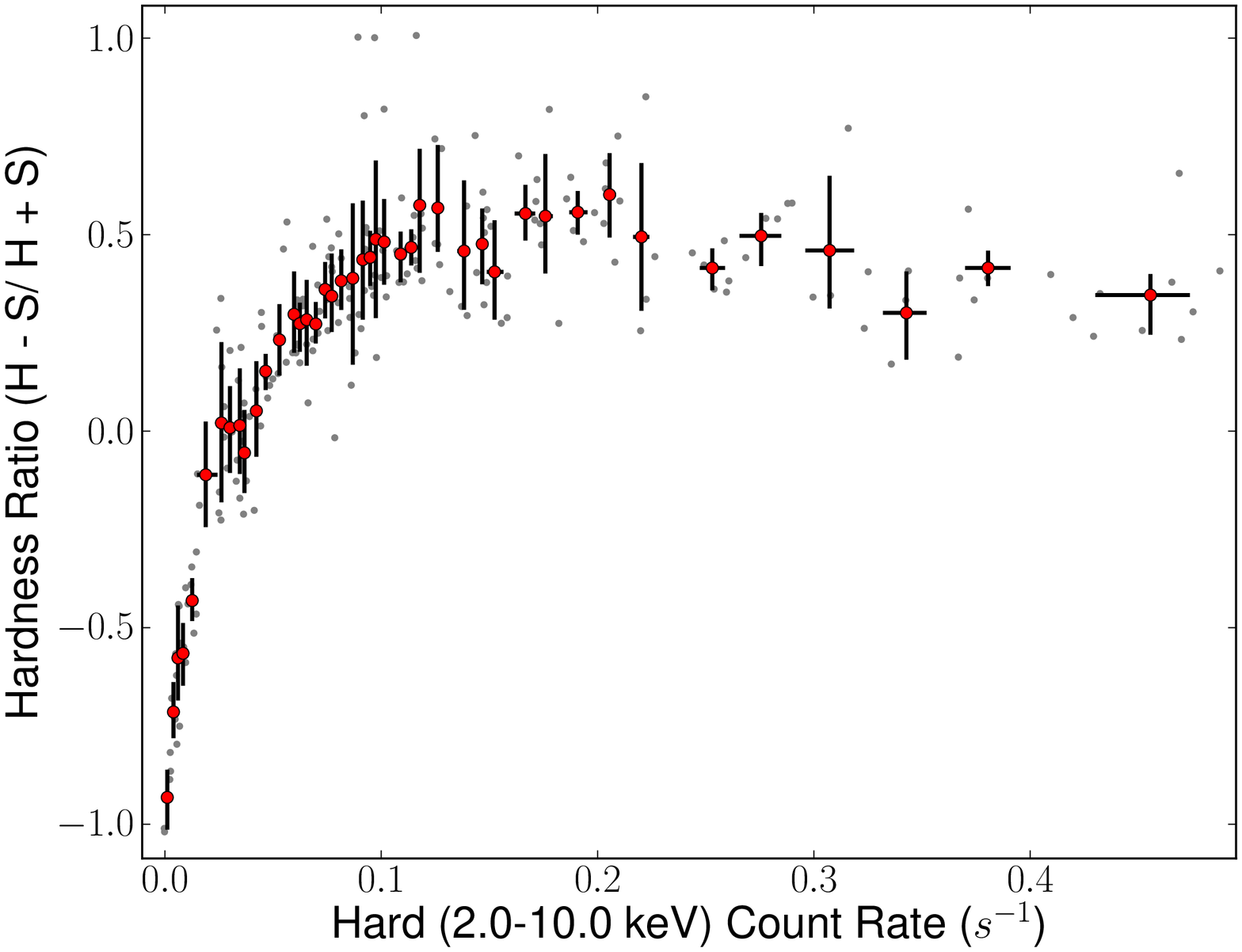}
	\includegraphics[width = \columnwidth,clip = true, trim = 0 0 0 0]{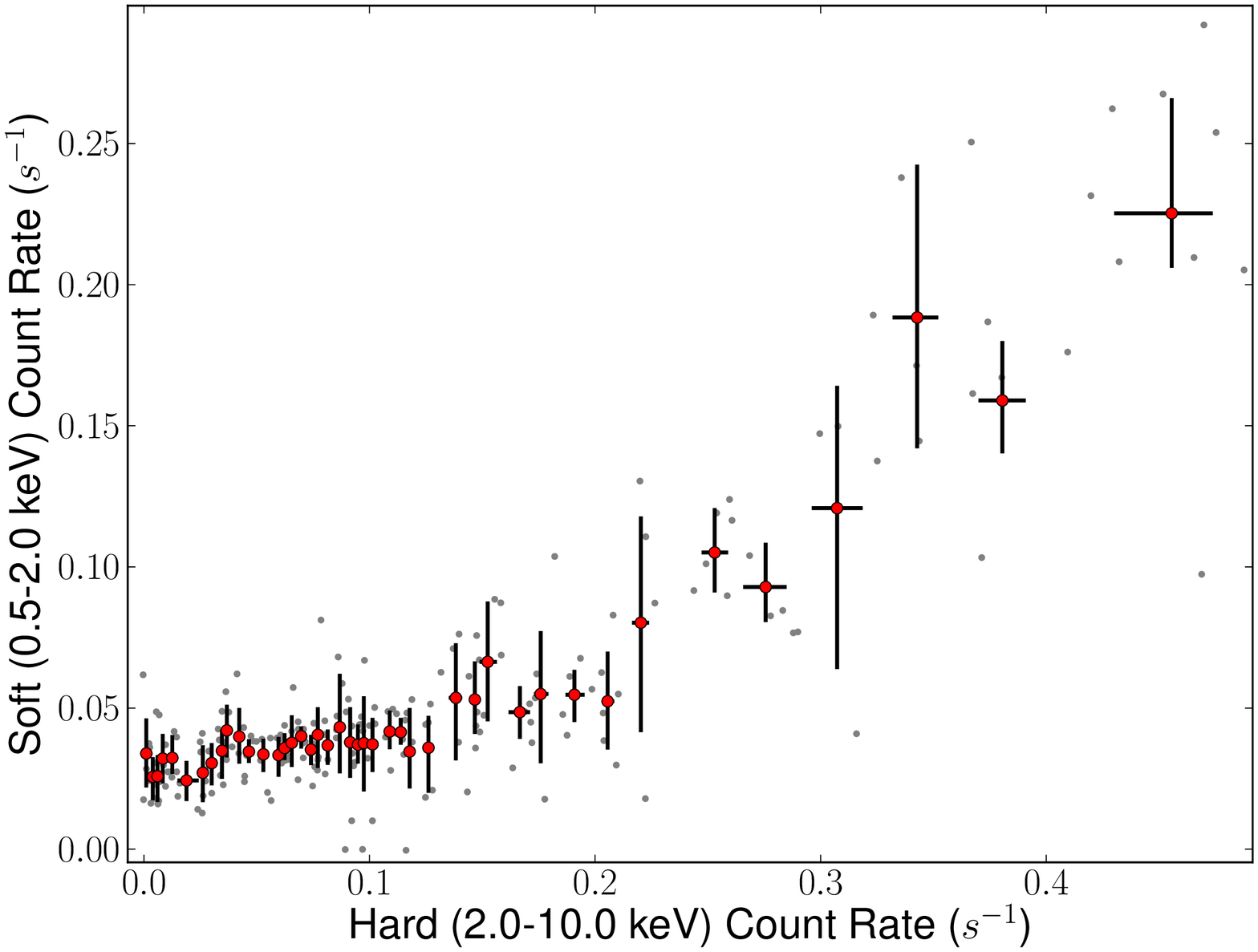}
	\caption{{\it Top:} Plot of the hard count rate against hardness ratio of NGC 1365. The scatter around the mean at a given count rate is low. {\it
Bottom:} Plot of the hard count rate against soft count rate of NGC 1365. In both plots, the data are binned such that each bin contains a minimum of 5 data points.
Errors in the hard count rate are the standard deviations of the distribution of points in each bin. The unbinned data are shown lightly behind the binned data. 
The typical fractional errors on the unbinned data are 0.16, 0.22 and 0.38 for the soft counts, hard counts and hardness ratio respectively.}
	\label{hardness}
\end{figure}

Plots of the hardness ratio against the hard count rate, and the hard count rate against the soft count rate are shown in Fig. \ref{hardness}. Whereas most previous
measurements of the photon index, $\Gamma$, concentrate on the $2.0-10.0$ keV energy band, here we look at spectral shape across a broader range of $0.5-10.0$ keV. As
lower energies are more sensitive to absorption, this energy range allows a more complete look at spectral changes due to absorbing material. In each case, hard emission
is defined as $2.0 - 10.0$ keV and soft emission as $0.5-2.0$ keV. The hardness ratio is defined as: 

\begin{equation}
\mathrm{Hardness\ Ratio} = \frac{\mathrm{Hard\ Cnt\ Rate} - \mathrm{Soft\ Cnt\ Rate}}{\mathrm{Hard\ Cnt\ Rate} + \mathrm{Soft\ Cnt\ Rate}}
\label{hardnessEqn}
\end{equation}
\hspace{0.5cm}
 
 Fig. \ref{hardness} shows the spectrum to be extremely soft at very low fluxes, but to become hard very rapidly with with increasing flux. Beyond
this sharp rise, still at a relatively low flux, the hardness decreases again more gradually with increasing flux, as often seen in Seyfert galaxies within the $2.0-10.0$
keV band \citep[e.g.][]{sobolewska,lamer}. The data display a relatively small amount of scatter about this general trend; for this reason, the shape of the spectrum can
be assumed to be approximately similar at a given flux level, independent of time. This implies that the system is behaving in approximately the same manner at each flux
level, irrespective of what state the system was in at an earlier time, allowing flux-binning of spectra to improve the signal-to-noise ratio.

\subsection{Spectral Modelling}

\begin{table*}
  \centering
  \footnotesize
  \begin{tabular}{p{0.01cm}p{4.6cm}p{4.15cm}>{\centering}p{0.8cm}>{\centering}p{1.0cm}>{\centering}p{1.1cm}>{\centering}p{1.5cm}>{\centering}p{0.3cm}p{0.6cm}}

	\hline 
	& Model 				    & Xspec description & Spectral index & Absorbing column & Ionisation & $\Gamma$ & $\chi^2_{Red}$ & DoF \\
	\hline 
	1 & Absorbed \& unabsorbed power laws & $\mathrm{powerlaw} + \mathrm{absori}\times\mathrm{powerlaw}$	& Fixed	& Free	& Free 	& 1.92 (fixed)	& 1.39	& 1329 \\
	2 & Absorbed \& unabsorbed power laws & $\mathrm{powerlaw} + \mathrm{absori}\times\mathrm{powerlaw}$	& Fixed	& Free 	& Tied 	& 1.92	(fixed) & 1.42	& 1339 \\
	3 & Absorbed \& unabsorbed power laws& $\mathrm{powerlaw} + \mathrm{absori}\times\mathrm{powerlaw}$& Fixed & Free & 0	& 1.92 (fixed) & 1.44	&
1340 \\
 	4 & Absorbed \& unabsorbed power laws & $\mathrm{powerlaw} + \mathrm{absori}\times\mathrm{powerlaw}$	& Fixed	& Tied	& Free  & 1.92	(fixed) & 2.28	& 1339 \\ 
	\hline
	5 & Absorbed \& unabsorbed power laws & $\mathrm{powerlaw} + \mathrm{absori}\times\mathrm{powerlaw}$	& Tied	& Free	& Free	& 1.47	(tied) & 1.21	& 1328 \\ 
	\hline
	6 & Absorbed \& unabsorbed power laws & $\mathrm{powerlaw} + \mathrm{absori}\times\mathrm{powerlaw}$	& Free	& Free	& Free	& 1.22 - 2.09	& 1.17	& 1318 \\
 	7 & Absorbed \& unabsorbed power laws & $\mathrm{powerlaw} + \mathrm{absori}\times\mathrm{powerlaw}$	& Free	& Free	& Tied	& 1.19 - 2.09	& 1.20	& 1328 \\
 	8 & Absorbed \& unabsorbed power laws & $\mathrm{powerlaw} + \mathrm{absori}\times\mathrm{powerlaw}$	& Free	& Tied	& Free	& 0.30 - 1.78	& 1.86	& 1328 \\ 
	\hline
	9 & Single, absorbed power	law   & $  \mathrm{absori}\times\mathrm{powerlaw}$			& Free	& Free 	& Free	& 0.61 -1.81	& 1.36	& 1328 \\
	10 & Single, absorbed power	law   & $  \mathrm{absori}\times\mathrm{powerlaw}$			& Free	& Free	& Tied 	& 0.58 -1.82	& 1.63	& 1328 \\ 
	11 & Single, absorbed power	law   & $ \mathrm{absori}\times\mathrm{powerlaw}$			& Free	& Tied	& Free	& 0 -1.91	& 1.42	& 1318 \\
 	\hline 
 	12 & Compton scattering		& $(\mathrm{powerlaw} + \mathrm{absori}\times\mathrm{powerlaw})\times\mathrm{cabs}$			& Fixed & Free &
Tied	& 1.92 (fixed) & 1.42	& 1339 \\
 	\hline 
	13 & Partial covering, fraction tied & $ \mathrm{pcfabs}\times\mathrm{powerlaw}$		& Fixed & Tied & n/a	& 1.92 (fixed) & 2.57	& 1351 \\
	14 & Partial covering, fraction tied & $ \mathrm{pcfabs}\times\mathrm{powerlaw}$		& Fixed & Free & n/a	& 1.92 (fixed) & 1.45	& 1351 \\
	15 & Partial covering, fraction free & $ \mathrm{pcfabs}\times\mathrm{powerlaw}$		& Fixed & Free & n/a	& 1.92 (fixed) & 1.45	& 1341 \\

 	\hline
	\end{tabular}
		
	\caption{Summary of the main components of each model fitted to the average spectra, showing the parameters which were fixed, tied or left free in each case, and
the value or range of values for the spectral index, $\Gamma$, which was fixed or best fitting. The reduced $\chi^2$ value and number of degrees of freedom (DoF) of the
best fit with each model is also shown. Each model also contains the fixed components of the best fit model to the diffuse gas around the nucleus
($\mathrm{tbabs}*\mathrm{apec}$) 
and the entire model is multiplied by the $\mathrm{wabs}$ model set to the galactic absorbing column.}
	\label{modelstable}
\end{table*}

\begin{table*}
  \centering
  \footnotesize
  \begin{tabular}{ c c c c c c c c }
  
	\hline 
	
Model 	 & $\Gamma$ 	     & $\mathrm{N_H}$ 	   	& $\mathrm{n_A}$  	& $\mathrm{n_U}$ 	        & $\xi$ 		    & T 		    &F \\
\hline 
1 & $1.92$ 		     & $9.48_{-1.72}^{+1.50}$  & $97.4_{-8.3}^{+8.2}$  & $3.53_{-0.42}^{+0.41}$  & $11.44_{-3.50}^{+5.57}$   & $10.00_{-3.57}^{+0.00\dagger}$ &$-$
\\
2 & $1.92$ 		     & $6.03_{-0.61}^{+0.85}$  & $90.2_{-7.0}^{+9.1}$  & $3.73_{-0.35}^{+0.37}$  & $0.41_{-0.36}^{+0.36}$    & $5.90_{-0.29}^{+4.10}$ & $-$ \\
3 & $1.92$ 		     & $5.82_{-0.69}^{+0.77}$  & $92.1_{-8.1}^{+8.7}$  & $3.80_{-0.36}^{+0.36}$  & $0$ 		     & $-$  			& $-$  \\
4*& $1.92$ 		     & $5.44$ 		   	& $84.6$ 		& $3.63$ 		    & $0.017$ 	 	        & $10.00$ 		  & $-$ \\
\hline
5 & $1.41_{-0.06}^{+0.05}$  & $8.33_{-2.18}^{+1.84}$  & $38.2_{-4.1}^{+4.4}$  & $0.395_{-0.46}^{+0.44}$  & $19.27_{-16.26}^{+12.63}$ & $10.00_{-2.56}^{+0.00\dagger}$& $-$
\\
\hline
6 & $1.22_{-0.18}^{+0.16}$  & $7.97_{-2.52}^{+2.08}$  & $25.8_{-8.1}^{+10.3}$ & $3.95_{-0.48}^{+0.46}$  & $30.99_{-30.99}^{+41.85}$ & $10.00_{-3.22}^{+0.00\dagger}$& $-$
\\
7 & $1.19_{-0.17}^{+0.16}$  & $5.01_{-0.86}^{+0.94}$  & $23.8_{-7.3}^{+9.4}$  & $4.20_{-0.40}^{+0.40}$  & $0.15_{-0.15}^{+0.65}$    & $3.40_{-2.85}^{+6.60}$ & $-$ \\
8 & $1.23_{-0.28}^{+0.12}$  & $8.84_{-0.62}^{+0.55}$  & $27.1_{-9.0}^{+7.6}$  & $4.01_{-0.46}^{+0.44}$  & $43.21_{-14.99}^{+65.38}$ & $7.60_{-2.03}^{+7.60}$ &$-$ \\
\hline
9 & $0.70_{-0.05}^{+0.07}$  & $3.91_{-0.79}^{+0.59}$  & $10.9_{-0.1}^{+0.1}$  & $-$ 			& $357.0_{-46.6}^{+55.5}$   & $3.04_{-0.38}^{+0.80}$	& $-$ \\
10 & $0.79_{-0.04}^{+0.04}$  & $3.84_{-0.75}^{+0.78}$  & $12.5_{-1.2}^{+1.3}$  & $-$ 			& $288.0_{-21.8}^{+23.0}$   & $2.70_{-0.32}^{+0.38}$	& $-$ \\
11 & $0.55_{-0.05}^{+0.08}$ & $1.29_{-0.07}^{+0.08}$  & $7.77_{-0.1}^{+0.1}$  & $-$ 			& $708.0_{-197.3}^{+334.0}$ & $0.28_{-0.05}^{+0.10}$	& $-$ \\
\hline
12 & $1.92$ 		     & $6.12_{-0.66}^{+0.78}$  & $91.2_{-7.6}^{+8.4}$  & $3.75_{-0.36}^{+0.36}$  & $0.39_{-0.35}^{+0.38}$ & $0.56_{-0.29}^{+9.54}$	& $-$ \\
\hline
13*& $1.92$ 		     & $4.51$		   	& $73.7$	   	& $-$			& $-$ 			     & $-$ & $0.952$ \\
14 & $1.92$ 		     & $7.50_{-0.78}^{+0.68}$  & $100.1_{-4.3}^{+9.3}$  & $-$ 			& $-$ 			     & $-$ &$0.963_{-0.006}^{+0.001}$ \\
15 & $1.92$ 		     & $7.57_{-0.85}^{+0.95}$   & $102.7_{-7.1}^{+5.7}$& $-$ 			& $-$ 			     & $-$ &$0.967_{-0.001}^{+0.02}$ \\

 	\hline
	\end{tabular}
		
	\caption{Typical parameter values for each model described in Table \ref{modelstable}, taken from the fit to the spectrum of the central flux-bin. 90\% errors
are
given where a parameter was free or tied. Errors for models with an asterisk (*) are not given, as the reduced $\chi^2$ of the fit was $>2$, preventing their calculation
in {\it XSPEC}. As the absorber temperature in models 5 and 6 reached the upper limit of the {\it absori} model, the positive errors on these values (indicated with
$\dagger$) are given as zero. These values are, however, unconstrained in the positive direction. The parameters are as follows:
$\Gamma$ - spectral index,
$\mathrm{N_H}$ - absorbing column 				($10^{22}$ cm$^{-2}$),
$\mathrm{n_A}$ - normalisation of the absorbed power law  	($10^{-4}$ keV$^{\ -1}$cm$^{-2}s^{-1}$ at $1$ keV),
$\mathrm{n_U}$ - normalisation of the unabsorbed power law 	($10^{-4}$ keV$^{\ -1}$cm$^{-2}$s$^{-1}$ at $1$ keV),
$\xi$ - ionisation state 					(L$_{5-300 \mathrm{keV}}$/N$_\mathrm{e}$R$^2$ - see \citealt{done92}),
T - absorber temperature					($10^{5}$ K),
F - covering fraction.}
	\label{valstable}
\end{table*}


The 190 separate time-resolved XRT spectra of NGC 1365 were divided into 11 flux bins and combined using the {\it HEADAS} tool `addspec' (see Fig. \ref{spectra} for a
sample of these summed spectra).  The bins were chosen such that each binned spectrum had both a minimum of 2000 total counts and a minimum width of 0.025 counts
s$^{-1}$ across the total range of hard count rates. This binning method ensured that each summed spectrum would possess a sufficient signal to noise ratio for accurate
spectral fitting, and that the flux bins were roughly evenly spaced across the hard flux range covered by the spectra. The energy channels of each of these summed spectra
were then grouped, using the {\it HEADAS} tool `grppha', such that each group contained a minimum of 15 counts. 

A variety of models were fitted to the spectra, using the {\it XSPEC} 12.7 analysis package \citep{arnaud}. The models were simultaneously fitted to all 11 summed
spectra, in order to discover the cause of variations in the shape of the spectrum with changing flux. In each case, parameters were either fixed at a single value
for all spectra, `tied' such that the parameter was allowed to vary but was the same for all 11 spectra, or left free to vary between the fits to individual spectra. 

 To account for the diffuse emission surrounding the nucleus of NGC 1365 (see Section \ref{obs}), the four {\it Chandra} spectra were fitted. These spectra were
composed of a minimum of 1300 counts. The spectral energy channels were grouped in the same way as the {\it Swift} spectra, such that each group contained a minimum of 15
counts.

As found previously by \citet{wang09}, the best fitting model is the {\it apec} model for collisionally-ionised diffuse gas combined with the {\it tbabs} model for
absorption by gas and dust. These models were used together with the  {\it wabs} model, to account for the known galactic absorption along the line of sight ($1.39 \times
10^{−20}$ cm$^{-2}$). A simultaneous fit to 4 regions, allowing N$_H$ to vary, but with all other parameters tied, gave a reduced $\chi^2$ of 0.96, with 197 degrees of
freedom. The absorbing column varied a little between regions. However, as a single value is required to fit the average contribution to the {\it Swift} nuclear spectrum,
we fitted all regions with the same $N_H$, giving a reduced $\chi^2$ of 1.46, with 200 degrees of freedom. In this model, the best fitting parameters were: an absorbing
column of $6.97^{-0.02}_{+0.02} \times 10^{20}$ cm$^{-2}$ in addition to galactic absorption, a gas temperature of $0.77^{-0.03}_{+0.03}$ keV, a metal abundance of
$0.069^{-0.010}_{+0.012}$ and a normalisation of the {\it apec} model of $5.51^{-0.63}_{+0.73} \times 10^{-4}$. The components of this model, with each of the parameters
fixed at the best fit values, were included in all of the subsequent fits of the {\it Swift} data. 

When modelling the resultant XRT background-subtracted nucleus, it was discovered that a single, unabsorbed power law of photon index $\sim 1.92$ (as found by
\citealt{risaliti13}), fitted the lowest flux observations very well, except for a small excess at higher energies ($> \sim 4.0$ keV). With increasing total flux, this
excess was seen both to increase in flux relative to the lower energy component and to expand to lower energies, such that the single power law model become increasingly
inadequate.

At the highest fluxes, it was found that a single power law would fit the data very well if absorbed by a partially ionised absorbing column. This model was, however,
still insufficient to give a good fit at intermediate fluxes, as these spectra had a comparable flux at both the low- and high-energy ends of the spectra.

Motivated by the steep unabsorbed spectrum, with its high energy excess, found at low fluxes, and the absorbed component found at the higher fluxes, a set of
two-component models were fitted to the data. These models consisted of two power laws, one of which was unabsorbed and the other absorbed, with the absorbing column and
ionisation parameter left as either free or tied parameters (see models 1-4 in Table \ref{modelstable}). These models were found to fit the data well in simultaneous
fits of the spectra from all flux ranges. The column of the absorbing material, and its ionisation parameter, were either tied or left to vary between individual spectra.
The matrix of the variations of each model, together with each goodness of fit, is given in Table \ref{modelstable}. Typical values of the parameters in each model are
given in Table \ref{valstable}.

In these two-component models, the two power laws had the same spectral index, $\Gamma$, fixed at the value found by \citet{risaliti13} using the very wide spectral
range allowed by {\it NuStar}. It was possible to obtain equally good fits at all fluxes using a single absorbed pivoting power law and better fits using two
pivoting power laws of the same spectral index (see models 6-11 in Table \ref{modelstable}). However, the best fitting models of this kind required a very large range of
$\Gamma$ (from $\sim 0.6$ to $> \sim 1.8$), and, very unusually, an extremely low value of $\Gamma$ at high fluxes. As described in Section
\ref{intro}, this is considered to be unphysical over the observed luminosity range.

In all models, the absorbed power law was modelled using the {\it absori} model for an ionised photoelectric absorber. The absorber
temperature was tied in all fits. A redshift of $5.569 \times 10^{-3}$ \citep{lavaux} and an iron abundance of 2.8 times solar abundance, as found by \citet{risaliti09},
were used in all fits. Galactic absorption of $1.39 \times 10^{−20}$ cm$^{-2}$ was also included in each model, using the {\it wabs} model for electromagnetic absorption
\citep{dickey90}. 

The effects of Compton Scattering on the fits were tested by adding the {\it cabs} model in {\it XSPEC} (see model 12 in Table \ref{modelstable}). The inclusion of
this
model was found to have very little effect on the best fit parameters, or the goodness of fit. This result is not unexpected, as the highest measured absorbing column  is
$1.0 \times 10^{24} $cm$^{-2}$, which is below the column required to be Compton thick ($1.5 \times 10^{24}$ cm$^{-2}$, see \citealt{malizia09}).

In model 3 of Table \ref{modelstable} the absorbing column is neutral, i.e. the ionisation state is fixed at zero. The best fit in this model is almost the same as those
in which the ionisation is free (models 1 and 2). Thus, although the value of the absorbing column has a large effect on our fits, the ionisation state is not well
constrained. 

Finally, a set of partial covering models were fitted to the data, as this type of model has previously been found to be successful in modelling changes over short
timescales \citep[e.g.][]{risaliti09} (see models 13-15 in Table \ref{modelstable}). The model ({\it pcfabs} in {\it XSPEC}) consists of a neutral absorber which covers a
fraction of the X-ray source, resulting in a spectrum composed of an absorbed and an unabsorbed component. We found that the data could not be well fit by a
model in which the absorbing column is constant between spectral fits, with only the covering fraction changing (model 13). If we allow the absorbing column to vary, we
obtain the same quality of fit whether we allow the covering fraction to vary (model 14) or not (model 15), as, when the fraction is allowed to vary, the same value is
derived at each flux level. Model 14 is physically very similar to our two component  model (model 2).

\subsection{Two-Component Spectral Variability}

\begin{figure}
	\includegraphics[width = \columnwidth, angle = 0,trim = 0 0 0 0,clip = true]{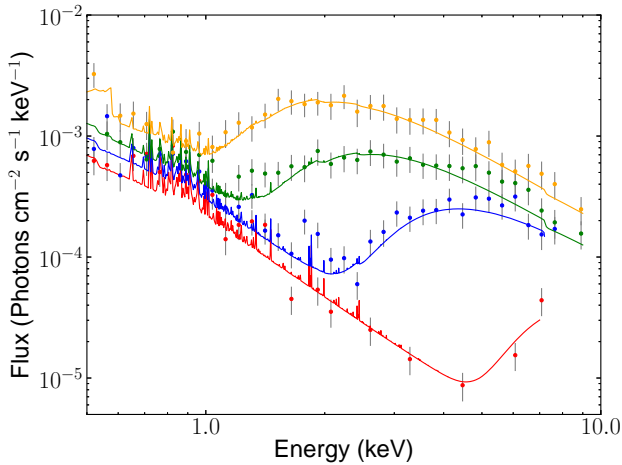}
	\caption{A sample of the set of unfolded, summed spectra produced by combining spectra in the same flux range. Alternate spectra are excluded, to prevent the
plot being crowded. The spectra are simultaneously fitted with the best fitting model, consisting of two power laws, one of which is absorbed by a partially ionised
absorbing column and one of which is not. Both power laws have a fixed spectral index of 1.92, as found by \citet{risaliti13}. The ionisation state of the absorbing
material is tied, but the absorbing column is allowed to vary between individual fits, producing the variation seen between spectra from different flux levels. The 
flux values are calculated by {\it XSPEC} using the model. The data are binned for clarity.}
	\label{spectra}
\end{figure} 

 \begin{figure}
	\includegraphics[width = \columnwidth, angle = 0,trim = 0 0 0 0,clip = true]{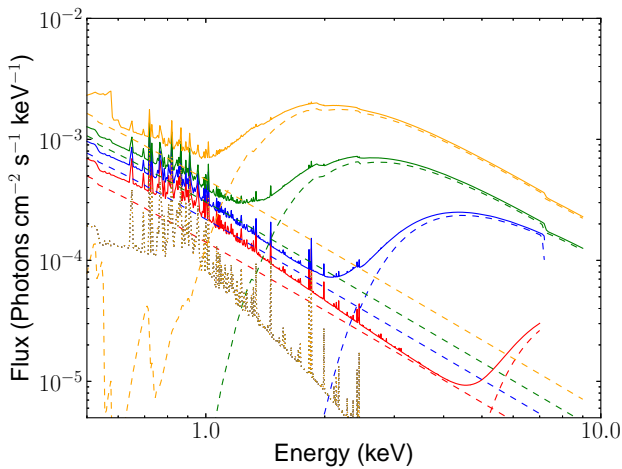}
	\caption{The sample of best fitting models shown in Fig. \ref{spectra} ({\it solid lines}), together with the components of each model - the unabsorbed power
	law and absorbed power law ({\it dashed lines}), and the spectrum of the underlying diffuse emission ({\it dotted line}).}
	\label{modelSpectra}
\end{figure} 
 

In all fits described here, we fix the underlying power law spectral index at the value given by \citet{risaliti13}. Allowing $\Gamma$ to vary does give better fits but,
as discussed above, the required range of $\Gamma$ is extremely large and almost certainly unphysical.

In the two-component models with a fixed $\Gamma$, leaving the ionisation state to vary, but tying the absorbing column, was
found to be insufficient to account for the degree of variation observed in the spectra, at any fixed absorbing column, for both a pivoting power law and the
two-component model, as can be seen from the $\chi^2$ values. 

Tying the ionisation state whilst leaving the absorbing column free to vary, however, gives good fits to the data. Significantly, the
$\chi^2$ value is very similar to that of the model in which both the ionisation state and absorbing column are both left free to vary, as the ionisation varies 
little in the best fit model. Whilst the ionisation state of the absorbing material undoubtedly changes, the data show both that large changes are not required
to give a good fit, and that large changes in the absorbing column are essential to account for the spectral variation observed regardless. Changes in ionisation alone
cannot account for these spectral changes observed. The two-component model with one absorbed and one unabsorbed power law, in which the absorbing column is left free to
vary between spectra whilst the ionisation state is kept constant, is therefore the simplest model with which the spectral variability seen in the data can be described.


In this model, there was a large range of absorbing columns, from  $1.69 \times 10^{22}$ to $10^{24}$ cm$^{-2}$. The temperature of the absorber was $9.5
\times 10^{5}$ K and the ionisation state of the absorber was 0.11. The normalisations of the two power laws varied by a factor of 3, giving a range of intrinsic
luminosities of approximately 1.1 to 3.5 ergs s$^{-1}$ for the 0.5-10.0 keV band.

\begin{figure}
	
	\includegraphics[width = \columnwidth,trim = 0 0 0 0, clip=true]{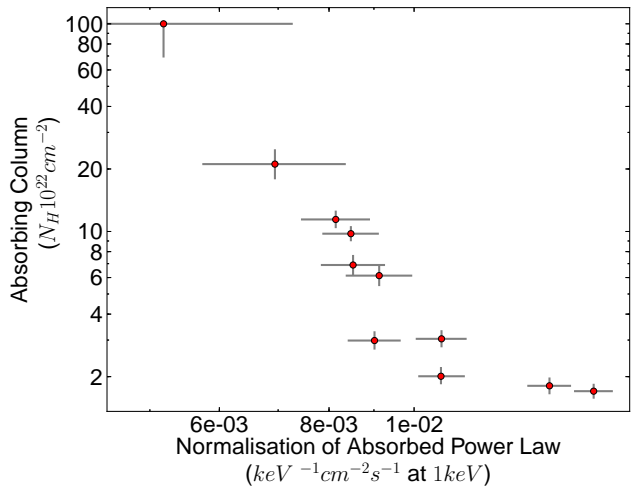}\hspace{1pt}
	\caption{Log plot of the normalisation parameter of the absorbed power law 
		against the column of the absorbing material in
		the model described above (model 2 in Table \ref{modelstable}). The normalisation of the power law is equivalent to the flux
		of this component before absorption, and therefore the X-ray luminosity of the source.}
	
	\label{nhplots}
\end{figure}

Fig. \ref{spectra} shows a sample of the flux-binned spectra fitted with this model. Fig. \ref{modelSpectra} shows how the two power law components vary as the flux
changes. The plot shows that the absorbing column of the absorbed power law decreases as the normalisation of the power law (i.e. the flux before absorption) increases.
These two parameters are plotted in Fig. \ref{nhplots}.  There is a strong decrease in the absorbing column as the normalisation increases (Spearman ranked
correlation coefficient $\rho = 0.98^{+0.02}_{-0.20}$), but the data are not of sufficient quality to precisely determ ne the form of the relationship between these
parameters. 

\begin{figure}
	
	\includegraphics[width = \columnwidth,trim = 0 0 10 0, clip=true]{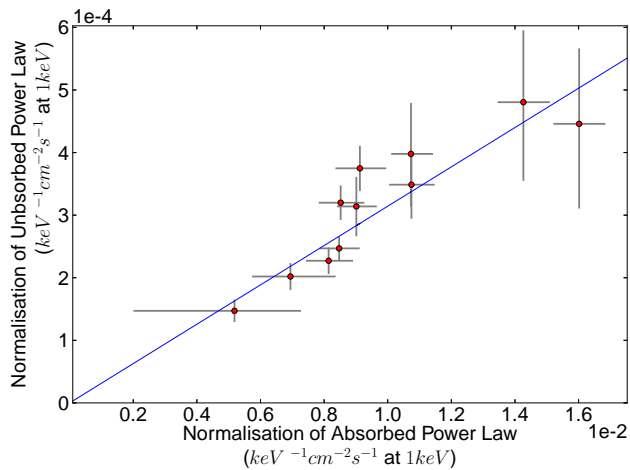}\hspace{1pt}
	\caption{Plot of the normalisation parameter of the absorbed power law 
		against the normalisation parameter of the unabsorbed power law in
		the model described above (model 2 in Table \ref{modelstable}). These normalisations are equivalent to the unabsorbed flux from
	        each component; the correlation between them implies the two components originate from the same source.  
	        The data are fitted with a linear model going through the origin, with a reduced $\chi^2$ of 1.50 for 10 degrees of freedom.}
	
	\label{normplots}
\end{figure}

In Fig. \ref{normplots} the normalisation of the unabsorbed power law is plotted against that of the absorbed power law. The normalisations can be
fitted well with a linear model, showing them to be correlated (r = 0.91). These observations confirm that there
is a real change in the underlying luminosity of the source and that the observed flux changes are not just due to changes in absorption. The $15-150$ keV {\it BAT}
lightcurve of NGC 1365, which undergoes far less absorption than the 0.5-10 keV band, confirms this intrinsic variability, showing variation of approximately a factor of
4.


\begin{figure}

	\includegraphics[width = \columnwidth,trim = 0 0 0 0, clip=true]{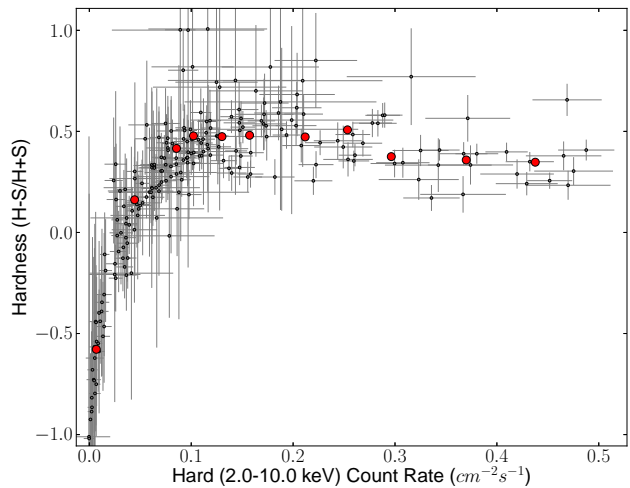}
	
	\caption{Plot of the hardness against the hard count rate of each of the 190 observed spectra ({\it small circles}), together with that of the model spectra
described above (model 2 in Table \ref{modelstable}) when fitted to each of the 11 summed, flux binned spectra ({\it large circles}).}
	\label{modelhardness}
	
\end{figure}

As a consistency check, the hard count rate and hardness ratio from each model fitted to the summed data are plotted over the original data as in Fig. \ref{hardness}
(Fig. \ref{modelhardness}). These plots show that the summed spectra also follow the trends shown by the individual spectra,
reinforcing the assumption that the spectrum is similar at a given flux level, independent of time.


A two component spectral model consisting of an unabsorbed scattered component and a more luminous absorbed component has previously been used to describe the single
epoch spectrum of NGC 4945 \citep{done96}. Broadly similar two-component spectra are also reported elsewhere  \citep[e.g. NGC4507,][]{braito}. Variations of
hardness ratio versus count rate similar to those shown here, in Fig. \ref{hardness}, have also been seen in the spectra of X-ray binary systems (Fig. 1 in
\citealt{kuulkers}) from RXTE data (2-13 keV). Although \citet{kuulkers} do not discuss flux-binned spectra as seen in our Fig. \ref{spectra}, they also favour a similar
two component model. These studies show that a two component model broadly similar to that which we describe here can, not infrequently, be used to parameterise the X-ray
spectra of both AGN and X-ray binary sources. However, the main contribution of this paper is to show, for the first time, that the complete, and very large, range of
spectral variability displayed by at least one AGN can be explained by systematic variation of the absorber, with the absorption varying inversely with luminosity. Below
we discuss a possible model to explain this behaviour. 


\section{Discussion}


\subsection{A Possible Relationship Between Source Flux and Column Density}
	
For the best-fitting two-component model described above (model 2 in Table \ref{modelstable}), Fig. \ref{nhplots} shows  the column of absorbing material to be inversely
related to the normalisation parameter of the absorbed power law, i.e. to the flux prior to absorption. As this component dominates the unabsorbed luminosity of the
source, the absorbing column is inversely proportional to the source luminosity. Whilst one might initially assume any reduction in absorption with increasing flux to be
due to increased ionisation, we have shown that models involving varying ionisation alone do not fit the data; models allowing variation of ionisation and absorbing
column simultaneously also require an inverse relationship between the absorbing column and the unabsorbed luminosity and do not require the ionisation to vary
significantly. Fits to spectra of NGC 4151 by \citet{lubinski10} at different flux levels show a similar reduction in the absorbing column with increasing flux, implying
that this relationship is not unique to NGC 1365. A physical mechanism is required to explain this relationship and a possible solution lies in an X-ray wind of absorbing
material rising from the accretion disc.

\subsection{AGN Wind Model}

X-ray absorption is often attributed to outflows, in particular to a disc wind (see e.g. \citealt{kaastra00}, \citealt{blustin05}, \citealt{tombesi}). In the AGN model
proposed by \citet{elvis}, absorbing material arises from a narrow range of
accretion disc radii in a biconical `wind'. \citet{nicastro} shows that an X-ray absorbing wind could originate from a narrow boundary region between the radiation
pressure-dominated and gas pressure-dominated regions of the accretion disc. In these models, a rise in accretion rate, which will give rise to an increase in X-ray
luminosity, naturally leads to an increase in the radii from which the wind arises through at least two mechanisms.  Firstly, a higher accretion rate leads to an
increase in disc temperature and hence an increase in the radii from which the wind arises.  Secondly, a higher accretion rate leads to an increased ionising UV flux from
the inner disc. When potential wind material at the disc surface is fully ionised, it is only subject to Compton scattering; the much more powerful line-driving
force \citep{proga99} is no longer applicable, hence this material would never be driven off as a wind. Thus, if the inclination is such that the observer views the
X-ray source through the inner edge of the wind, an increase in accretion rate would move the wind outwards such that the observer would now be viewing the source through
a lower absorbing column.  An additional geometric factor which would reduce absorbing column is that, in the \citet{elvis} model, an increase in X-ray luminosity causes
the opening angle of the wind to increase due to the increased radiation pressure. The above affects will thus naturally lead to the inverse relationship between
absorbing column and luminosity which we see. Moving the absorbing wind to larger radii also means that the line of sight would go through the less dense, more highly
ionised front part of the wind, as we see in our fits.

The low ionisation states found in our models suggest that the material involved is not associated with the highly-ionised, high-velocity outflows suggested by
\citet{risaliti05a}, which is too ionised to cause the observed X-ray absorption. We therefore suggest that the absorbing material lies in a region which is further out
than this highly ionised material, as part of a stratified wind (\citealt{tombesi}, \citealt{elvis}).

Depending on which mechanism dominates in pushing the wind to larger radii with increasing accretion rate, we expect different lags between the change in absorbing column
and the change in luminosity which could lead to hysteresis and scatter in the X-ray luminosity / hardness relationships.

If the outward movement of the wind launch radii is caused by a change in disc temperature due to inwardly propagating accretion rate fluctuations, then the
X-ray luminosity will lag by the viscous travel time of those fluctuations to either the inner edge of the disc where the seed photons are mainly produced and/or to the
X-ray emitting corona itself. For typical wind launch radii of a few hundred R$_{\mathrm{g}}$ \citep{higginbottom}, this timescale would be of order weeks to months. 

If the outward movement of the wind is dominated by an increase in ionising UV photons from the inner edge of the disc, which also dominate the X-ray seed photon flux,
then variations in the X-ray luminosity will lead changes in the absorbing column slightly, by the difference between the light travel time from the UV region to the wind
and to the corona respectively. This difference is likely to be small (hours). Alternatively, if the wind moves outwards in response to increased UV flux, but the X-ray
luminosity rises in response to increased accretion rate rather than increased seed photon flux, then the X-ray luminosity will lag by the difference between the viscous
propagation time from the UV to X-ray emitting regions and the light travel time from the UV emitting to wind launch radii. This difference is also likely to be small
(1-2 days).

Our observations do not show a great deal of scatter in the hardness-count rate diagram, indicating that any lags are short. We therefore favour a mechanism by which the
wind launch radii are pushed outwards mainly by a rise in ionising UV flux, rather than by a rise in local disc temperature. Regardless of mechanism by which these
outflows are driven, it is clear from our data that a geometrical response to changes in luminosity is necessary in the absorber in order to explain the variability in
the spectra.

Finally, we note that, in all models, the wind is expected to be clumpy, meaning short timescale variations of the absorbing column, independent of the unabsorbed
luminosity, should also occur, which would also add scatter to the relationship between X-ray luminosity and absorbing column. The spectral variations reported previously
on short timescales are probably mainly the result of fitting variations induced as clumps pass over the line of sight.

\section{Conclusions}

{\it Swift} X-ray observations of NGC 1365 over a period of 6 years show a large amount of spectral variability. These variations are best explained by a two-component
model
consisting of an unabsorbed power law and a more luminous absorbed power law; for both components, the spectral index was fixed. The normalisations of the two power laws
vary together. The absorbing column of the absorbed power law varies inversely with its luminosity, an effect which is not simply due to increased ionisation. This effect
can be simply explained by viewing through the edge of a wind whose launch radius varies inversely with increasing accretion rate. The unabsorbed power law could be
explained, as in the standard \citet{elvis} wind model, as the scattered component from the far side of the wind.

\section*{Acknowledgments}

SDC thanks the STFC for support under a studentship and IMcH thanks the STFC for support via grant ST/G003084/1. We thank Christian Knigge and James Matthews for useful
discussions.


\label{lastpage}

\end{document}